# Metal-Insulator and Superconductor-Insulator Transitions in Correlated Electron Systems

Masatoshi Imada[1], Fakher F. Assaad[2], Hirokazu Tsunetsugu[3] and Yukitoshi Motome[4]

1) Institute for Solid State Physics, University of Tokyo, Roppongi, Minato-ku, Tokyo 106-8666, Japan
e-mail address imada@issp.u-tokyo.ac.jp
2) Institut für Theoretische Physik, Teilinstitut III, Universität Stuttgart, Pfaffenwaldring 57, D-70550 Stuttgart, Germany
3) Institute of Applied Physics, University of Tsukuba, Tsukuba, Ibaraki 305-8573, Japan
4) Department of Physics, Tokyo Institute of Technology, Oh-okayama, Meguro-ku, Tokyo 152-8551, Japan

**Abstract**
Quantum transitions between the Mott insulator and metals by controlling filling in two-dimensional square lattice are characterized by a large dynamical exponent $z = 4$ where the origin of unusual metallic properties near the Mott insulator are ascribed to the proximity of the transition. The scaling near the transition indicates the formation of flat dispersion area due to singular momentum dependence of the single-particle renormalization. The flat dispersion controls critical properties of the Mott transition. An instability of the flat dispersion to the $d$-wave superconducting order is discussed. We also discuss a case of the Mott transition for a model of Mn perovskite compounds with orbital degeneracy where orbital correlation length shows critical divergence toward the metal-insulator transition.

## 1 Introduction

Metal-insulator transitions (MIT) driven by strong correlation effects are called Mott transitions and are a subject of recent intensive studies in two- and three-dimensional systems [1]. For one-dimensional (1D) systems, the Mott transitions are relatively well understood where rigorous results can be obtained for some nontrivial cases. In infinite-dimensional systems where the dynamical mean field theory becomes exact, the MIT can also be studied in a controlled way. However, in 1D systems, the Fermi level (and the Fermi surface) is represented by only two points while in the infinite-dimensional systems, the single-particle self-energy is site-diagonal and has no wavenumber dependence. Therefore, in both cases, charge excitations near the Fermi level are not allowed to have momentum dependence along the Fermi surface. On the contrary, as we see below, the momentum dependence of the single-particle renormalization may have singular dependence along the Fermi surface in finite-dimensional systems such as in two and three dimensions and therefore would show quite different features from one- and infinite- dimensional systems.

In two or higher dimensional systems, there exists no evidence to exclude the Fermi liquid phase as the only one stable fixed point of ordinary metals. However, this is not necessarily an isotropic metal. When charge excitations in a part of the Fermi surface have slow dynamics, they are liable to couple strongly to other degrees of freedom ( such as spin fluctuations) thus yielding even slower dynamics. This synergetics may selfconsistently generate singular momentum dependence of the single-particle renormalization, where at a particular region (or points) on the Fermi surface, the Fermi liquid description breaks down. This particular region may be characterized by flat dispersion with mass enhancement due to the real part of the selfenergy as well as by strong damping due to the imaginary part.

Metallic phase near the Mott insulator is known to show various unusual properties in terms of standard metals. However, in any experimental systems clear phase boundary is not observed between the paramagnetic metallic phase near the Mott insulator and more or less standard metals observed far from the correlated insulator. This strongly suggests that the unusual properties are not due to the appearance of a new phase with adiabatic discontinuity from the weakly correlated metals, but due to a proximity of the MIT. Therefore, it is important to understand critical properties of the MIT. Metal-insulator transitions are typical examples of quantum phase transitions which takes place at zero temperature by changing a parameter to control quantum fluctuations rather than thermal fluctuations. In this paper, we discuss that the unusual properties can be understood from novelty of the quantum criticality at the transition to the Mott insulator [2, 3, 4]. In quantum phase transitions, dynamical fluctuations play roles even in high spatial dimensions. These are correctly treated in the "dynamical mean field theory" which provides exact result if spatial dimensionality is infinite and spatial fluctuations are suppressed [5]. In real materials, however, quantum fluctuations with strong interaction effects are most conspicuous in systems with low-dimensional anisotropy. In this case, fluctuations near quantum phase transition points are enhanced and appear as combined effects of strong spatial and dynamical quantum fluctuations.

The Drude weight defined from the coefficient of the $\delta$-function in the frequency-dependent conductivity at $T = 0$ as

$$\sigma(\omega) = D\delta(\omega) + \sigma_{\text{reg}}(\omega) \tag{1}$$

and the charge compressibility $\kappa$ defined below are the two important and relevant quantities to describe the Mott transition. The Mott insulator has two basic properties: one is insulating property and the other, incompressibility. The Drude weight and the charge compressibility are both zero in the Mott insulating phase while both nonzero in metallic phases. Therefore, they are totally nonanalytic at the transition point. The nonanalyticities in these two quantities are also common in the transition to the band insulator, while they do not have singularities in case of the Anderson localization transition.

We here discuss in more detail the important difference between the Anderson transition and the MIT to the Mott insulator. In the Anderson transition, it is well known that the singularity at the transition point appears in the diffusion constant or the relaxation time of the carrier $\tau$ (inverse of the imaginary part of the single-particle self-energy) but neither in the carrier effective mass $m^*$, nor in the density of states $N(\omega)$ etc. [6] Therefore the DC conductivity $\propto n\tau/m^*$ is a good probe to determine the character of the transition because it is directly proportional to the relaxation time. In contrast, in the transition to the Mott insulator, the singularity may not appear in the relaxation time of the carrier but appears in the ratio of the carrier density $n$ to the effective mass $m^*$. The DC conductivity in the ideal case (without disorder at $T = 0$) is always infinite in the metallic phase and cannot be a good probe in the ideal condition (it can be a probe to some extent if the disorder or thermal effects play some role) while the Drude weight $D \propto n/m^*$ and the charge compressibility $\kappa$ discussed here become relevant quantities with singularities.

We note that the quantities which explicitly depend on the relaxation time arising from thermal or disorder effects have to be treated carefully. As we see below it is possible that the MIT is controlled by a particular part of flat dispersion on the Fermi surface. However, the relaxation-time-dependent quantities such as DC transport properties can be "contaminated" by the relaxation time and the real criticality could be obscured because the contribution from the part of flat dispersion could be suppressed due to shorter relaxation time as compared to the other part.

We also note the difference between the Mott transition and magnetic quantum phase transitions which take place within metals. In the latter case, self-consistent renormalization theory was developed [7, 8] where, at $T=0$, a Gaussian treatment from the paramagnetic phase and the Hartree-Fock-RPA description are justified above the upper critical dimension. There the anomalous character of metals is mainly ascribed again to the anomaly in the relaxation time although a weak effect on the renormalization factor is expected.

In §2, we review results on novel universality class of two-dimensional MIT driven by correlation effects. The scaling theory with the universality class characterized by the dynamical exponent $z=4$ offers a unified description for the unusual properties of metal near the Mott insulator. Numerical results on the charge compressibility, spin correlations, dynamical conductivity in the metallic phase as well as the localization length in the insulator are consistently understood from this quantum criticality. The strong momentum dependence of the renormalization around $(\pi, 0)$ spot is suggested as the basis of the scaling description. Superconductor-insulator transitions have been studied motivated from instability of anomalous metallic states associated with this novel universality class of the MIT. The results are summarized in §3. MITs with orbital degeneracy is the subject of §4.

## 2 Metal-Insulator Transition in Two Dimensions

### 2.1 Numerical Results

The Mott insulator and metallic states near the MIT have been intensively studied subject by numerical approach. Here, we see that the 2D single-band Hubbard and $t$-$J$ models show consistency with the scaling theory [3, 4]. It is discussed later that the single-particle excitation of the 2D Hubbard model around $(\pm\pi, 0)$ and $(0, \pm\pi)$ in the momentum space is renormalized to form a flat quartic dispersion. This singular renormalization effect determines the universality class of the transition between the Mott insulator and metals and constitutes a basis for the scaling theory. This circumstance is a unique feature of the Mott transition in finite-dimensional systems in contrast with one and infinite dimensions.

Before the clarification of the full momentum dependence, the singular momentum dependence was suggested from the "momentum integrated" quantities. The first signature was observed in the charge compressibility. The charge compressibility $\kappa$ or charge susceptibility $\chi_c$ are defined as $\chi_c = n^2 \kappa = \partial n / \partial \mu$ where $n$ is the electron density and $\mu$ is the chemical potential. The charge susceptibility vanishes in the Mott insulating phase because of the incompressibility. However, quantum Monte Carlo results on 2D systems show that doped systems become more and more compressible near the MIT in the metallic side [9, 10]. The scaling plot of $\chi_c$ shows singular dependence on the doping concentration $\delta$ in the form
$$\chi_c \propto 1/|\delta|^p \tag{2}$$
with $p = 1$ for $\delta \neq 0$ [9, 10, 11]. Similar is also observed in the 2D $t$-$J$ model [12, 13]. When the charge susceptibility is singularly divergent as a power of $\delta$ for $\delta \to +0$ in 2D, the single-particle description of low-energy excitations has to show the divergence of the effective mass of relevant particle. This is in contrast with the usual MIT between metals and the band insulator, where the number of carriers vanishes.

The second signature of unusual renormalization was observed in the measurement of the localization length in the insulating phase [14]. A large numerical advantage to observe the transition from the insulator side is that we can avoid the negative sign problem known in the quantum Monte Carlo method. The insulator undergoes a transition to a metal when the chemical potential $\mu$ approaches the critical point $\mu_c$ from the region of the charge gap. In the insulating phase, the single-particle Green function defined as

$$\mathcal{G}(r,\tau) = \langle Tc(r,\tau)c^\dagger(0,0)\rangle \tag{3}$$

provides the localization length $\xi_l$ defined by

$$\mathcal{G}(r,\omega=\mu) \equiv \int_0^\infty \mathcal{G}(r,\tau,\mu)d\tau \sim e^{-r/\xi_l}. \tag{4}$$

$\xi_l$ may be regarded as the localization length of the wavefunction of virtually created state at the chemical potential inside the gap. The scaling of $\xi_l$ near $\mu_c$ calculated with the above method by Assaad and Imada [14] for the two-dimenional Hubbard model shows

$$\begin{aligned} \xi_l &\sim |\mu - \mu_c|^{-\nu} \\ \nu &= 0.26 \pm 0.05. \end{aligned} \tag{5}$$

The obtained $\nu$ is consistent with $\nu = 1/4$.

The third signature is seen in the Drude weight. The Drude weight was calculated for the $t$-$J$ model by the exact diagonalization[15]. The doping concentration dependence in Fig. 1 shows that the total kinetic energy $\langle -F\rangle$ is proportional to $\delta$ while the Drude weight $D$ vanishes much faster consistently with $D \propto \delta^2$. It should be noted that this is in sharp contrast with a naive expectation that the carrier density is proportional to $\delta$, where the Drude weight in the Drude theory is proportional to $n/m^*$ and should be linear in the carrier density $n$ divided by the carrier effective mass $m^*$. As we discuss in §2.3, the frequency dependence of the conductivity in numerical results also suggests that the most part of the conductivity weight in the sum rule is exhausted in a rather universal line shape of incoherent conductivity $\sigma_{\text{reg}}(\omega)$ and is consistent with a quick collapse of the Drude part with decreasing doping.

The above signatures observed in the charge compressibility, localization length and dynamical conductivity are hard to understand when isotropic renormalization of charge excitations is assumed along the large Fermi surface which satisfies the Luttinger theorem.

## 2.2 Scaling Theory of Mott Transition

The numerical results on the charge compressibility and spin correlations in 2D have inspired extensive studies to clarify the nature of the Mott transition in low-dimensional systems [2, 3, 4] . Because the MIT is controlled by quantum fluctuations, we may take the control parameters such as the electron chemical potential $\mu$ or the bandwidth $t$. Using this control parameter, the distance from the critical point is measured by $\Delta$. The control parameter can either be the chemical potential to control the filling or the bandwidth (or the interaction). The scaling theory assumes the existence of single characteristic length scale $\xi$ which diverges as $|\Delta| \to 0$ and a single characteristic frequency scale $\Omega$ which vanishes as $|\Delta| \to 0$. The correlation length $\xi$ is assumed to follow $\xi \sim |\Delta|^{-\nu}$ as $\Delta \to 0$, which defines the correlation length exponent $\nu$. The frequency scale $\Omega$ is determined from the quantum dynamics of the system independently of the length scale. The dynamical exponent $z$ determines how $\Omega$ vanishes in relation to $1/\xi$ as $\Omega \sim \xi^{-z} \sim |\Delta|^{z\nu}$. The hyperscaling asserts the homogeneity in the singular part of the free energy density $f_s$ in $d+1$ dimensional path integral in terms of arbitrary length-scale transformation parameter $b$:

$$f_s(\Delta) \sim b^{-(d+z)} f_s(b^{1/\nu}\Delta) \sim \Delta^{\nu(d+z)}. \tag{6}$$

When the inverse temperature $\beta$ and the linear dimension of the system size $L$ are both large but finite, finite-size scaling function $\mathcal{F}$ is expected to hold as

$$f_s(\Delta) \sim \Delta^{\nu(d+z)} \mathcal{F}(\xi/L, \xi^z/\beta) \tag{7}$$

in the combination of non-dimensional arguments.

Combining expressions in the path integral formalism for the Drude weight and the compressibility, respectively, with the finite-size scaling form derived from the hyperscaling, one obtains useful scaling forms for physical quantities [3, 4, 16]. The scaling of $D$ is obtained:

$$D \propto \Delta^\zeta \tag{8}$$

with $\zeta = \nu(d + z - 2)$, while

$$\delta \propto \Delta^{-\alpha + \nu z} \tag{9}$$

and

$$\kappa \propto \Delta^{-\alpha} \tag{10}$$

are obtained with $\alpha = \nu(z - d)$, where $\delta$ denotes the carrier concentration, namely, the density measured from the Mott insulating phase.

In the insulating phase, the charge excitation gap $E_g$ and the localization length $\xi_l$ defined in (4) should be determined from $E_g \propto |\Delta|^{z\nu}$ and

$$\xi_l \propto |\Delta|^{-\nu}, \tag{11}$$

where we have used the fact that $\nu$ and $z$ are identical in both sides of the transition point. In the metallic side, the scaling of the specific heat $C = \beta^2 \partial^2 (\ln Z)/\partial \beta^2$ is given from (7):

$$C = \Delta^{\nu(d+z)} T \frac{\partial^2}{\partial T^2} \mathcal{F}(\xi/L = 0, \xi^z T). \tag{12}$$

At the critical point, the specific heat is given from the $\Delta$ independent term as $C \propto T^{d/z}$. If $C = \gamma T$ is satisfied at low temperatures in the metallic phase, the coefficient $\gamma$ follows

$$\gamma \propto \Delta^{\nu(d-z)}. \tag{13}$$

Another interesting quantity is the coherence temperature $T_F$ below which the electron motion becomes quantum mechanical and degenerate. It is clear that $T_F$ has to approach zero as $\Delta \to 0$ in a continuous transition. In case of the Fermi liquid, $T_F$ is nothing but the Fermi temperature. The existence of single characteristic energy scale with singularity at $\Delta = 0$ leads to the scaling of $T_F$ in the form

$$T_F \propto \Delta^{\nu z}. \tag{14}$$

When the control parameter $\Delta$ is the chemical potential $\mu$ measured from the critical point $\mu_c$, it represents the filling control MIT. In this case, it is possible to derive a useful scaling relation. Because the doping concentration is $\delta = -\frac{\partial f_s}{\partial \mu} = -\frac{\partial f_s}{\partial \Delta}$, it scales as

$$\delta \sim \Delta^{\nu(d+z)-1} \tag{15}$$

From the comparison of Eq.(15) and Eq.(9), $\nu z = 1$ is derived. The characteristic length scale is then

$$\xi \sim \delta^{-1/d}, \tag{16}$$

which is the length scale of the "mean hole distance". The above scaling description is known to be valid for the MIT to the band insulator, all the MIT in 1D [17] and the Anderson transitions [18].

## 2.3 Universality Class

When we assume the above scaling theory with Eq.(10), the quantum Monte Carlo result (2) implies that there exists a new universality class $z = 1/\nu = 4$ [2, 3]. As we see above in (5) and (11), the localization length also suggests that $\nu = 1/4$. These are independent estimates of the exponent and a check for the consistency of the scaling theory. On the contrary, these two results can be explained neither by the Hartree-Fock approximation nor by the $d = \infty$ results. Indeed, neither the Hartree-Fock nor the $d = \infty$ results satisfy the hyperscaling assumption (6). For example, the Hartree-Fock approximation predicts $\nu = 1/2$.

The large dynamical exponent $z$ leads to unusual suppression of coherence in the metallic phase in various aspects. The scaling theory predicts that the Drude weight scales as $D \propto \delta^{1+\frac{z-2}{d}}$. As is well known, $D \propto \delta$ is satisfied in the usual MIT, which is consistent with the scaling theory when $z = 2$. When the universality class is characterized by $z = 1/\nu > 2$, $D$ is suppressed stronger than $\delta$-linear dependence at small $\delta$. When $z = 4$ is satisfied in 2D, $D \propto \delta^2$ is predicted. This is one of the indications for the unusual suppression of coherence in this universality class. Numerical results indeed showed consistency with $D \propto \delta^2$ as we discussed in §2.1 [15].

From the sum rule, the $\omega$-integrated conductivity gives the averaged kinetic energy

$$-\langle F \rangle = \int_0^\infty \sigma(\omega)d\omega. \tag{17}$$

In the strong coupling limit as in the $t$-$J$ model, the total kinetic energy $\langle F \rangle$ is expected to be proportional to $\delta$ as observed in Fig.1a. Then the Drude part $D \propto \delta^{1+\frac{z-2}{d}}$ becomes negligibly small at small $\delta$ in the total weight $-\langle F \rangle \propto \delta$ if $z > 2$. This means most of the total weight $-\langle F \rangle$ is exhausted in the incoherent part $\sigma_{\text{reg}}(\omega)$ in (1). In the scaling theory, the form of $\sigma_{\text{reg}}(\omega)$ is not specified and indeed it may depend on details of systems including the interband transition. However, for the intraband contribution, $\sigma(\omega)$ is expected to follow $\sigma(\omega) \sim (1 - e^{-\beta\omega})/\omega$ at very high temperatures $T$ larger than the bare bandwidth $t_B$. In fact the conductivity is given by

$$\sigma(\omega) = \frac{1 - e^{-\beta\omega}}{\omega} C(\omega) \tag{18}$$

with the current correlation function $C(\omega)$ defined by

$$C(\omega) \equiv \int_{-\infty}^\infty dt\, e^{i\omega t} \langle j(0)j(t) \rangle \simeq \frac{\sigma_0/\tau}{-i\omega + \frac{1}{\tau}}, \tag{19}$$

where if the carrier dynamics is incoherent, $C(\omega)$ is expected to have a broad featureless structure due to rapid decay of current correlation in time $\tau \sim 1/t_B$. When the temperature is lowered, this incoherent part $\sim (1 - e^{-\beta\omega})/\omega$ is in general transferred to the Drude part $\sim \frac{1}{-i\omega+\gamma}$ below the coherence temperature. In the usual band-insulator-metal transition, this transfer can take place for the dominant part of the weight because $-\langle F \rangle$ and $D$ both may be proportional to $\delta$. Therefore, the incoherent part is totally reconstructed to the Drude part with decreasing temperature below $T_F$. However, for $z > 2$, this transfer takes place only in a tiny part of the total weight because the relative weight of the Drude part to the whole weight is $\delta^{(z-2)/d}$ which vanishes as $\delta \to 0$ for $z > 2$. Even at $T = 0$, the major part of the conductivity weight must be exhausted in the incoherent part. Near the transition point, since we have no reason to have a dramatic change of the form of the incoherent part $\sigma_{\text{reg}}(\omega)$ at any temperatures from the scaling point of view, the optical conductivity more or less follows the form

$$\sigma_{\text{reg}}(\omega) \sim C \frac{1 - e^{-\beta\omega}}{\omega} \tag{20}$$

even for $t_B > \omega$ and $t_B > T$ where we have neglected model dependent feature such as the interband transition. This "intraband" incoherent weight may be proportional to $C \propto \delta$ in the strong coupling limit. Such dominance of the incoherent part with the scaling (20) is consistent with the numerical results of 2D systems[19], which lends further support for $z > 2$ in 2D. On the contrary, the numerical results in 1D by Stephan and Horsch [20] support that the incoherent part is small at low temperatures. This may be due to $z = 2$ in 1D because the majority of the weight seems to be exhausted in the Drude weight.

Another indication for the suppression of coherence may be seen in the coherence temperature $T_F$. The scaling of $T_F$ is given by $T_F \propto \delta^{z/d}$. Because the standard MIT is characterized by $T_{F0} \propto \delta^{2/d}$, the relative suppression $T_F/T_{F0}$ is proportional to $\delta^{\frac{z-2}{d}}$. When $z = 4$ as suggested by numerical results in 2D, we obtain $T_F \propto \delta^2$ in 2D in contrast with $T_{F0} \propto \delta$ for $z = 2$.

In the Mott insulating phase, entropy coming from the spin degrees of freedom is released essentially to zero when the antiferromagnetic order exists. The entropy is also released when the spin excitation has a gap as in the spin gap phase. Even without such clear phase change, the growth of short-ranged correlation progressively releases the entropy with decreasing temperature. When carriers are doped, an additional entropy due to the charge degrees of freedom is introduced in proportion to $\delta$. This additional entropy is assigned not solely to the charge degrees of freedom but also to the spin entropy through their mutual coupling which destroys the antiferromagnetic long-range order. This additional entropy $\propto \delta$ has to be released below the coherence temperature $T_F$. Therefore, a natural consequence is that if $T$-linear specific heat characterizes the degenerate (coherent) temperature region, the coefficient $\gamma$ should be given by $\gamma \sim \delta/T_F \sim \delta^{1-\frac{z}{d}}$. This is indeed the scaling law we obtained in (13) and (15). From this heuristic argument, it turns out that the metallic phase near the Mott insulator is characterized by large residual entropy at low temperatures which is also related with the suppression of the antiferromagnetic correlation as compared to the insulating phase. It may also be said that the anomalous suppression of the charge coherence at small $\delta$ is caused by short-ranged antiferromagnetic correlations which scatters carriers incoherently.

The strong coupling of spin and charge seems to yield the same scaling even for the spin correlation. The numerical result suggests that the antiferromagnetic transition takes place more or less simultaneously with the Mott transition [9]. The equal time spin structure factor $S(Q)$ at its maximum shows the scaling $S(Q) \propto 1/\delta$ from which the magnetic correlation length $\xi_{AFM}$ is scaled by $\delta^{-1/2}$. The incommensurate wavevector $Q$ approaches $(\pi, \pi)$ as $\delta \to 0$. This is the same scaling as (16) and implies that the criticality of the antiferromagnetic transition is also involved in the same universality.

## 2.4 Flat Dispersion and Strong Momentum Dependence

To clarify the origin of the large dynamical exponent, $z = 4$, it is necessary to study explicit momentum dependence of the charge excitations. The realization of the hyperscaling must be the consequence of the fact that some small number of singular points in the momentum space appear and the MIT is controlled by the charge excitations around those points. If the charge excitations near the whole $d - 1$ dimensional Fermi surface would contribute equally and isotropically, it would not satisfy the hyperscaling at all because it would introduce a length scale of the Fermi wavenumber which does not vanish at the MIT point. This additional length scale would invalidte the assumption (6). Furthermore, the dynamical exponent $z = 4$ directly has to lead to the appearance of a $k^4$ dispersion around those singular points because by definition $z$ is nothing but connecting characteristic wavenumber and energy scales. If a flat dispersion such as $k^4$ appears at some points on the Fermi surface, the holes will be doped predominantly to this region because of this flatness and this flat region will determine the character of the MIT.

To understand whether the above reasoning works, the single-particle spectral function $A(k,\omega)$ was calculated and analyzed in detail. $A(k,\omega)$ of a doped single hole to the Mott insulator phase of the Hubbard model has peak structures which disperses in the momentum space. Figure 2 shows this dispersion around $(\pi,0)$ and clearly indicates that the flat dispersion does appear around $(\pi,0)$ and $(0,\pi)$, while other part of the Fermi level does not show such flatness. The numerical result further indicates that the dispersion around $(\pi,0)$ and $(0,\pi)$ is well fit by $k^4$ dispersion. These points have originally the van-Hove singularity in the simple Hubbard model. However, the observed $k^4$ dispersion is far beyond the expectation from the van-Hove singularity and the consequence of strong correlation effects. We further note that this $z=4$ universality class also holds for the case with finite next-nearest-neighbor hopping $t'$ where the van-Hove singularity does not meet the Fermi level at the MIT point [10]. This gives us the microscopic basis to understand the anomalous feature of the MIT.

The dispersion of a doped hole in the Mott insulating state does not necessarily determine the criticality of the transition because the rigid coherent band picture is not always guaranteed in the process of further doping. However, the existence of the flat dispersion is commonly observed at low doping concentration in angle resolved photoemission (ARPES) data of high-Tc cuprates [21, 22]. The presence of this flat dispersion is also consistent with numerical results of the Hubbard models at finite doping [23]. More remarkable is the following: The ARPES data for the insulating phase of the cuprates appears to show a deeper level around $(\pi,0)$ than the level around $(\pi/2,\pi/2)$ which is relatively closer to the Fermi level as is consistent with the case with the next nearest neighbor transfer $t'$ [24]. If the rigid band picture would be correct, this implies that the state around $(\pi/2,\pi/2)$ would be first doped with holes rather than $(\pi,0)$ region. However, in the underdoped region, the data indicate that the state around $(\pi,0)$ seems to quickly emerge from the incoherent tail of $A(k,\omega)$ at the Fermi level and forms a flat dispersion, which indicates the breakdown of the rigid band picture and a universal dominance of the flat dispersion excitations upon doping. This is also consistent with the numerical compressibility data with $t'$ discussed above. The flat dispersion could be obscured by its incoherent nature and hard to observe incoherent tails in $A(k,\omega)$ by the strong damping.

The dominance of the flat dispersion around $(\pi,0)$ on low energy excitations should be taken with care because it has some degeneracy with the excitations around $(\pi/2,\pi/2)$. It may have an effect on the DC transport properties, although the DC properties are not good probes to understand the Mott transition as we discussed in the introduction. The ARPES data of the underdoped cuprates show a gradual formation of fragmentary "Fermi surface" first around $(\pi/2,\pi/2)$ in the normal state. This may have two reasons. One is that the carrier relaxation time around $(\pi/2,\pi/2)$ is much longer and the other is that the region around $(\pi,0)$ is under the influence of the formation of pseudogap. The low-energy excitation around $(\pi,0)$ is dominated by the paired singlet formation due to the instability of the flat dispersion as we will discuss in the next section. The paired singlet formed in the pseudogap region seems to be still incoherent above $T_c$ [25, 26]. Although the MIT is governed by the excitations around $(\pi,0)$, the DC transport may have substantial contribution from the part around $(\pi/2,\pi/2)$. This is speculated from the fragmentary Fermi surface as well as from a rather large and sensitive increase of the residual resistivity upon Zn doping around 1% in the cuprates while relatively insensitive effect of Zn on the pseudogap behavior [27, 28]. This implies that the Zn doping sensitively induces the localization of carriers around $(\pi/2,\pi/2)$, while gapped singlet excitations formed from $(\pi,0)$ excitations is not seriously influenced by such tiny amount of Zn. The DC charge transport is strongly affected from $(\pi/2,\pi/2)$ excitations while other quantities including spin excitations are dominated by the driving excitations of MIT near $(\pi,0)$, which makes the physical properties as if spin and charge degrees of freedom were "separated". However, the seeming separation has to be understood after considering the strong momentum dependence carefully.

## 3   Superconductor-Insulator Transition

In the Mott insulator, because the single-particle process is suppressed due to the charge gap, we have to consider the two-particle process expressed by the superexchange interaction. This has led to the Heisenberg model for the effective Hamiltonian of the Mott insulating phase. The Hubbard model would be sufficient even for the Mott insulating phase if its low energy excitations could be precisely considered. However, the Heisenberg model offers a better and easier way to extract physics of spin excitations and properties at low temperatures in the strong coupling regime. Similarly to this circumstance in the insulating case, even in the metallic region, such suppressions of coherence discussed above for the single-particle process make it useful and helpful to consider the two-particle process explicitly. This is particularly true in the flat-dispersion part of the momentum space. In contrast with the insulating phase, the two-particle processes in metals contain an effective pair hopping term and a term described by

$$\mathcal{H}_W = -t_W \sum_i [\sum_{\delta\sigma}(c_{i\sigma}^\dagger c_{i+\delta\sigma} + c_{i+\delta\sigma}^\dagger c_{i\sigma})]^2 \qquad (21)$$

becomes relevant [29, 30, 31]. Then the Hamiltonian containing the term $\mathcal{H}_W$ added to the Hubbard model, called the $t$-$U$-$W$ model is expected to be a useful effective Hamiltonian to understand low temperature properties near the Mott insulating phase. It was argued that $\mathcal{H}_W$ plays the same role as the term proportional to $t^2/U$ in the strong coupling expansion of the Hubbard model which contains the superexchange term as well as the so-called three-site term [30]. It is not clear at the moment whether the bare Hubbard model implicitly contains sufficiently large effective coupling of $\mathcal{H}_W$ to make it relevant at low energy scale or some additional elements are needed. However, by including a small amount of the $\mathcal{H}_W$ term explicitly, it allows to study how the Hubbard hamiltonian becomes unstable to the two-particle process and what type of symmetry breaking or ordering are expected as its consequence. In fact the $t$-$U$-$W$ model is the first hamiltonian which has made possible to study the $d$-wave superconductor-Mott insulator transition under a controlled numerical treatment.

We note that, to treat the relevance of the $\mathcal{H}_W$ term under a proper condition, it would be necessary to first extract precisely the strong momentum dependence of the single-particle renormalization in the Hubbard model. Without clarifying the formation of the flat dispersion caused by many-body effects described in the previous section, the real role and relevance of the two-particle term cannot be understood well. Mean field analyses of the Hubbard as well as the $t$-$J$ models in the literature have not properly treated this important point by neglecting the singular momentum dependence of the renormalization.

When the two-particle process becomes relevant, the coherence is not suppressed because the universality class of the two-particle transfer is given by $z = 1/\nu = 2$. The universality class $z = 2$ is numerically observed in the exponent of the localization length in the insulating side of the $t$-$U$-$W$ model [31] and the doping concentration dependence of the Drude weight of the $t$-$J$-$W$ model [15]. Because of small $z$, it always becomes more relevant than the single-particle transfer for small $\delta$. The alteration of the universality class from $z = 4$ for the MIT to $z = 2$ for the superconductor-insulator transition shows an instability of metals near the Mott insulator to the superconducting pairing or in more general an instability of $z = 4$ universality phase to some type of symmetry-broken state.

Below, we summarize how the Mott insulator to superconductor transition takes place and how the $d$-wave superconducting state appears in the 2D $t$-$U$-$W$ model. This has been studied through large-scale quantum Monte Carlo calculation [29, 30, 31]. The $t$-$U$-$W$ model appears to show a Mott insulator to superconductor transition even when the filling is fixed at half filling. With the increase of $W$, the antiferromagnetic long-range order seen in the pure Hubbard model continuously decreases and at the critical amplitude of $W = W_c$, the order is destroyed. At the same $W$, the insulating phase appears to undergo a transition to a superconductor with $d_{x^2-y^2}$ symmetry.

A remarkable property in the superconducting phase at half filling is that the antiferromagnetic correlation is extremely compatible. The equal-time antiferromagnetic correlation decays very slowly with a power law at long distance as $\sim 1/r^\alpha$ with $\alpha \sim 1$ or slightly larger. Reflecting this compatibility, the real part of the staggered susceptibility $\chi(q = (\pi, \pi), \omega = 0)$ appears to diverge with lowering temperature to the limit $T \to 0$. A clearer understanding is obtained in the dynamical structure factor $S(q, \omega)$ (or equivalently in the imaginary part of the dynamical susceptibility $\text{Im}\chi(q, \omega)$). From this quantity, it turns out that the antiferromagnetic correlation is dynamical and has a strong peak at a finite frequency with presumable divergent weight in the thermodynamic limit. This peak frequency is comparable to the pairing gap amplitude. Because of the appearance of the superconducting phase, the antiferromagnetic correlation is suppressed below the pairing energy scale. However very slow power-law decay of equal-time antiferromagnetic correlation implies that $S(q = (\pi, \pi), \omega)$ has divergent weight somewhere at finite frequency. To reconcile the antiferromagnetic and superconducting correlations, the frequency of the antiferromagnetic correlation is pushed out from low-frequency to the frequency above the pairing gap. When the temperature is lowered at $W > W_c$ from high temperatures, $S(q = (\pi, \pi), \omega)$ first shows growth of a broad peak around $\omega = 0$. This appears to be an immatured state which does not differentiate the antiferromagnetic and singlet correlations. This is reminiscent of the SO(5) scenario with approximate symmetry of superconducting and antiferromagnetic state [32], although the approximate symmetry in this model has not been examined in detail yet. With further lowering of the temperature, the peak position shifts to a finite frequency and is sharpened. In this temperature range pseudogap is formed where $\text{Im}\chi(q = (\pi, \pi), \omega)$ decreases at low $\omega$. At lower temperatures, the peak at finite frequency becomes even sharper.

## 4  Mott Transition with Orbital Degeneracy

Orbital degeneracy plays important roles in addition to the spin degeneracy in a wide class of materials at the MIT [1]. It provides another origin of residual entropy near the transition point and may offer richer structure of the transition. Effects of orbital degeneracy was recently examined by projector quantum Monte Carlo method at zero temperature for a model of Mn perovskite compounds with degeneracy of two $e_g$ orbitals, $d_{x^2-y^2}$, and $d_{3z^2-r^2}$, under the condition of complete spin polarization and two-dimensional configuration [33]. In Mn perovskite compounds, the Mott insulator is in the $d^4$ configuration where one of two $e_g$ orbitals are occupied. Through the MIT, the spin configuration is always completely ferromagnetic within a plane mainly due to a strong Hund's rule coupling to high-spin $t_{2g}$ electrons. However, the charge transport is strongly incoherent with very small Drude weight in the metallic region [34]. This suggests a crucial role of orbital degeneracy (and presumably also the lattice degrees of freedom coupled to it) for the incoherent charge dynamics. The spin polarized but orbitally degenerate model may be a minimal model to capture this circumstance.

The Hamiltonian of this model is defined as

$$\begin{aligned}\mathcal{H} &= -\sum_{\langle ij\rangle,\nu,\nu'} t_{i,j,\nu,\nu'}(c_{i\nu}^\dagger c_{j\nu'} + h.c.) \\ &\quad + U\sum_{i,\nu\neq\nu'}(n_{i\nu}-\frac{1}{2})(n_{i\nu'}-\frac{1}{2}) - \mu\sum_{i\nu} n_{i\nu},\end{aligned} \quad (22)$$

where the creation (annihilation) of the single-band electron at site $i$ with orbital $\nu$ is denoted by $c_{i\nu}^\dagger(c_{i\nu})$ with $n_{i\nu}$ being the number operator $n_{i\nu} \equiv c_{i\nu}^\dagger c_{i\nu}$. An important difference from the ordinary Hubbard model is that the transfer has an anisotropy with dependence on the $x^2 - y^2$ and $3z^2 - r^2$ orbitals, where we label the orbital $x^2 - y^2$ and $3z^2 - r^2$ as 1 and 2, respectively below. In 2D configuration, the nearest-neighbor transfer is scaled by a single parameter $t$ as $t_{11} = \frac{3}{4}t$, $t_{22} = \frac{1}{4}t$, $t_{12} = t_{21} = \pm\frac{\sqrt{3}}{4}t$, where, in $\pm$, $+(-)$ is for the transfer to the $y(x)$ direction. In the absence of $U$, the diagonalized two bands have dispersions both with the bandwidth $4t$, where one dispersion is obtained by a parallel shift of $2t$ from the other.

The present numerical result is summarized as follows. When $U$ is increased, the Mott gap $\Delta_c$ opens at "half filling", namely at $\langle n \rangle = 1$. However, the opening of the gap is substantially slower than the case of the ordinary single-band Hubbard model (with spin). For example, $\Delta_c \simeq 0.1$ at $U/t = 4$ is compared with $\Delta_c \simeq 0.66$ for the ordinary Hubbard model at $U/t = 4$. At $U/t = 3$, $\Delta_c$ is not distinguished from zero within the error bar. Although the opening of the Mott gap is slow within this model, the insulating state at $\langle n \rangle = 1$ is stabilized if we introduce the Jahn Teller distortion. A realistic Jahn-Teller coupling for the Mn compounds has stabilized the Jahn-Teller distorted insulating state with staggered order of $3x^2 - r^2$ and $3y^2 - r^2$ orbitals. Another important observation is that realistic amplitudes of the Jahn-Teller coupling in the absence of $U$ is far insufficient in stabilizing the Jahn-Teller distortion with a realistic stabilization energy of several hundred K. The Mott insulating state in the experimental situation may well be resulted from synergy of $U$ and the Jahn-Teller coupling. In the model (22) with 2D configuration, the long-ranged orbital order increases at $\langle n \rangle = 1$ with increasing $U$. For example the staggered orbital polarization at $U = 4t$ is around 0.3. However, the orbital order appears to be destroyed immediately upon doping. In the metallic region $n \neq 1$, the short-ranged correlation of the staggered orbital order is critically enhanced as $\langle n \rangle \to 1$ similarly to the ordinary 2D single-band Hubbard model. For large $U$, there seems to exist a critical region near $\delta \equiv 1 - n = 0$ where $T(Q) \propto 1/\delta$ as the same as the spin structure factor of the ordinary Hubbard model. Here $T(Q)$ is the equal-time structure factor at $(\pi,\pi)$ for the orbital correlation. This critical enhancement of the orbital correlation strongly suggests that the orbital correlation in this Mn model may play a similar role to spins in the Hubbard model in realizing highly incoherent charge dynamics.

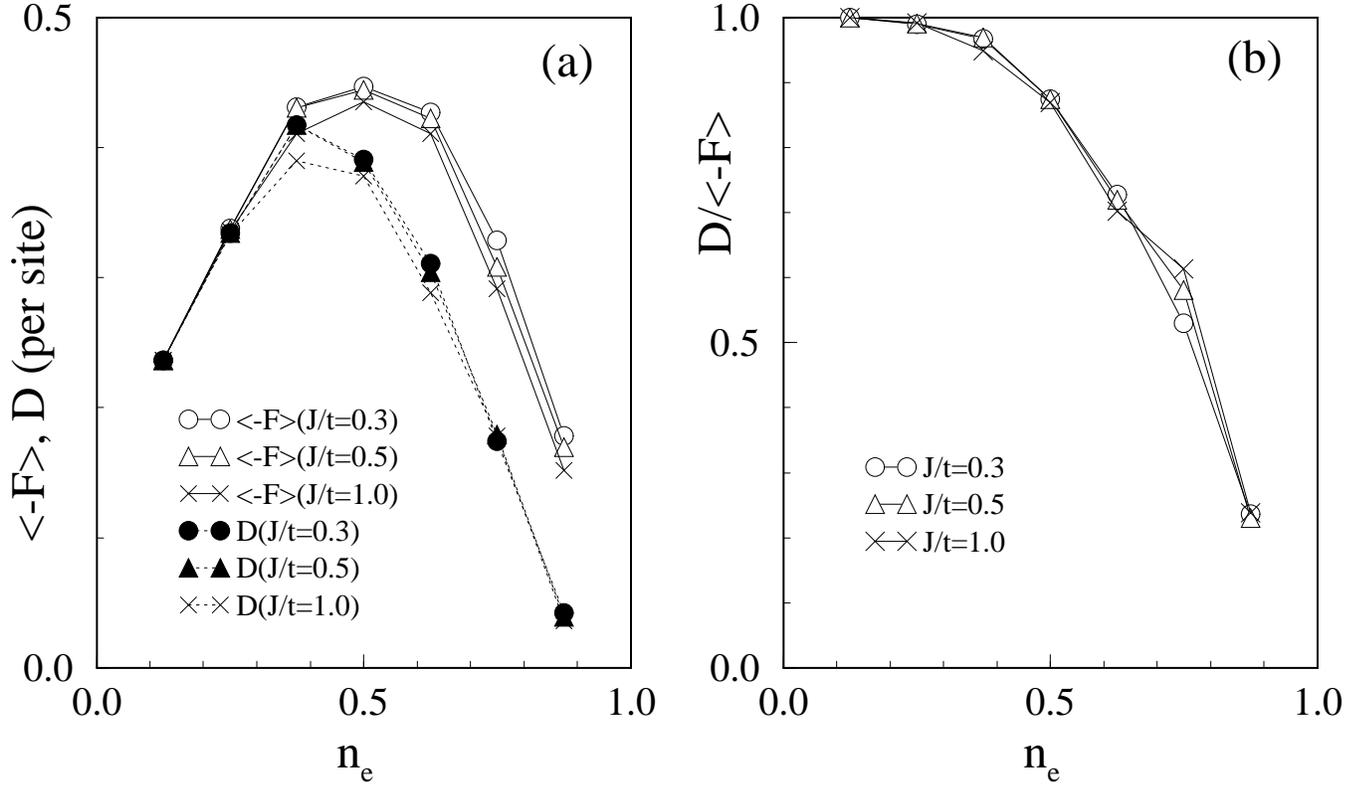

Figure 1: (a) Drude weight $D$ and total weight $\langle -F \rangle$ in the optical conductivity of the 2D $t$-$J$ model with 4×4 sites at zero temperature. (b) The ratio of the two weights $D/\langle -F \rangle$.

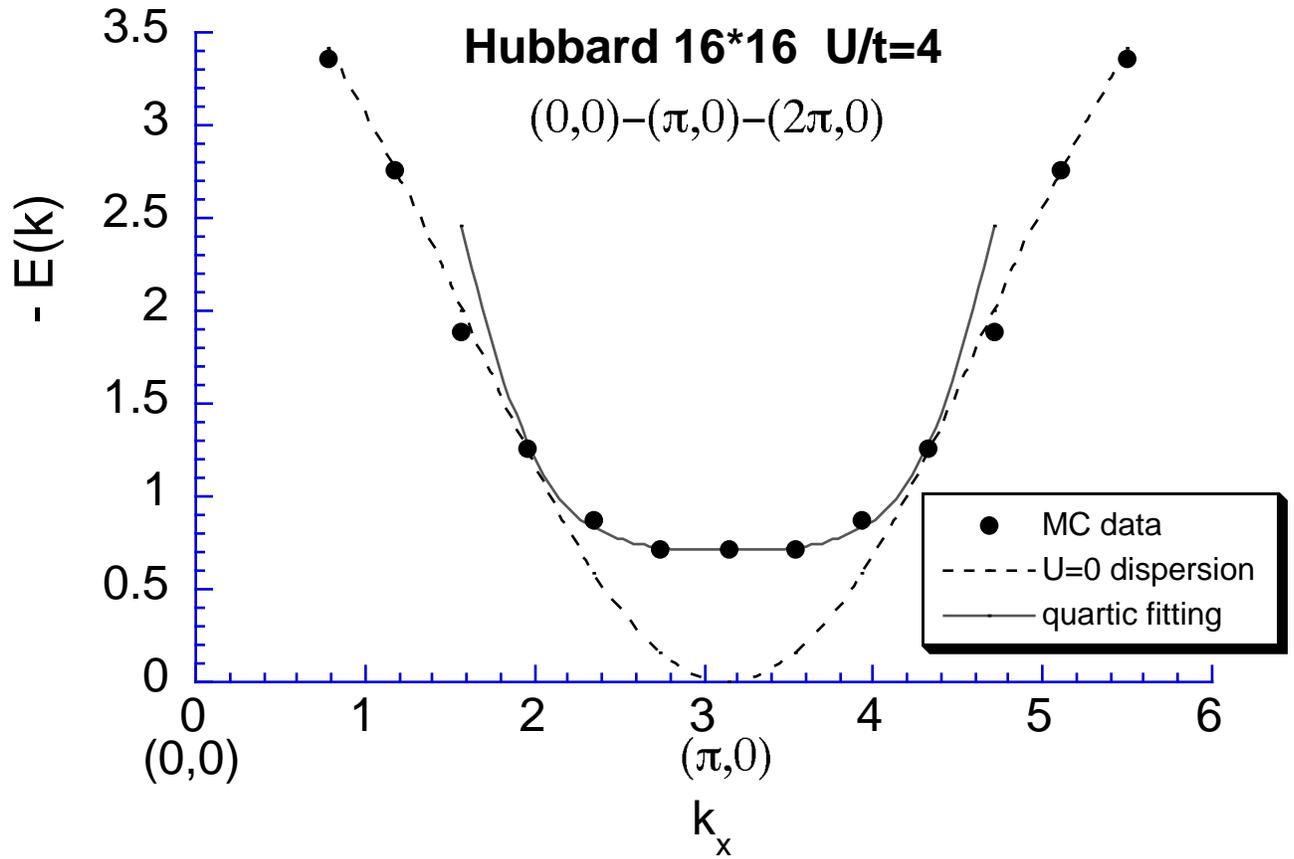

Figure 2: The dispersion of the Hubbard model along the $k_x$ axis, from $(0,0)$ through $(\pi,0)$ till $(2\pi,0)$ obtained from the peak position of the single-particle spectral weight $A(k,\omega)$ for the 16 by 16 lattice at zero temperature. Black circles are the Monte carlo data at $U/t = 4$.